\documentclass{aa}
\usepackage{natbib}
\usepackage{xspace}
\usepackage{amssymb}
\usepackage{amsmath}
\usepackage{graphicx}
\usepackage{rotating}
\usepackage{natbib}

\begin{document}

  \title{Planetesimal formation near the snow line in MRI-driven
    turbulent protoplanetary disks}

  \titlerunning{Particle growth around the snow line}
  
  \authorrunning{Brauer,
    Henning \and
    Dullemond}
  
  \author{F.~Brauer \and 
    Th.~Henning  \and
    C.P.~Dullemond}

  \date{\today} 

  \institute{Max-Planck-Institut f\"ur Astronomie, K\"onigstuhl 17,
    D--69117 Heidelberg, Germany; e--mail: brauer@mpia.de}

  \abstract{The formation of planetesimals in protoplanetary disks due to
    collisional sticking of smaller dust aggregates has to face at least two
    severe obstacles, namely the rapid loss of material due to radial inward
    drift and particle fragmentation due to destructive collisions. We present
    a scenario to circumvent these two hurdles. Our dust evolution model
    involves two main mechanisms. First, we consider a disk with a dead
    zone. In an almost laminar region close to the midplane, the relative
    velocities of the turbulent particles are comparatively small, which
    decreases the probability of destructive particle collisions. Second,
    turbulence is not the only source of violent relative particle velocities,
    because high radial drift speeds can also lead to boulder fragmentation.
    For this reason, we focus additionally on the snow line. Evaporation
    fronts can be associated with gas pressure maxima in which radial drift
    basically vanishes. This implies that particle fragmentation becomes even
    less likely. Our simulation results suggest that particles can overcome
    the fragmentation barrier. We find that boulders of several $10^2$~m can
    form within only a few thousand years.}

\maketitle


\section{Introduction}

The process of planetary formation is widely accepted to be initiated in
circumstellar disks by the collisional coagulation of dust grains to larger
objects \citep{h06,natta07}. Evidence of grain growth to particle sizes beyond
those typically observed in the interstellar medium is provided by
mid-infrared spectroscopy of disks around young stars
\citep{bouwman01,boekelaebe03,apai04,kessler07,sic}. Analyses of millimeter
interferometry data of disks implies that large populations of dust grains
exist with particle radii up to several millimeters \citep{Tes03,Wil05,Rod06}.

Models of dust particle growth in protoplanetary disks predict coagulation,
which supports the idea of planetesimal formation by particle hit and stick
mechanisms. Numerical calculations support the formation of cm-sized particles
within a few $10^3$ orbital timescales in the inner parts of the disk
\citep{Wei80}. After particles have grown to larger sizes, they can also be
affected by other processes described by alternative planet formation
scenarios. For example, large grains experience vertical settling, which
involves the formation of a dense midplane dust layer \citep{garlin2004}. This
layer can be affected by gravitational instability
\citep{GolWar73,Wei79,YouShu02,schr04,garlin2004} -- an issue that is still
highly debated. Due to the overwhelming amount of observational data for
protoplanetary disks available, models of dust particle growth have attracted
more attention \citep{tan05,DD05,nomura06,allesio06,ormel07_2,ciesla07}. All
of these particle evolution models can explain observational disk features
such as, for example, the group I/II disk classification \citep{meeus01,dd04_}
or the disappearance of infrared excess in the spectra of disks with ages
higher than a few Myrs \citep{hai01,carp05}, although, there are still many
open observational riddles that these models will hopefully answer in the near
future.

Models of protoplanetary disks do not only help to interpret observational
data; they also identify the serious obstacles to planetesimal formation
\citep{you04,domppv07}. One of these obstacles is particle fragmentation. Due
to the high particle relative velocities in disks that can reach 100~m/s
\citep{Wei77,ormel07}, particle collisions lead to particle destruction
instead of particle growth \citep{BluWur99,brauer2,anners08}. Depending on the
disk model, dust particle growth is inhibited significantly by high-speed
impacts of particles around a meter size. Even for times as long as 1~Myr,
solid particles are unable to overcome the fragmentation barrier and the
formation of planetesimals as precursors for Earth-like or Jovian planets,
therefore, poses a major problem. Despite this theoretically predicted
inability for planetesimals to form, the orbital decay of dust particles can,
nevertheless, reproduce the observed evolution in the spectral energy
distribution \citep{hai01,carp05}. The presence of a growth barrier is,
therefore, not in contradiction with observations.

Several mechanisms in protoplanetary disks can produce planetesimal
formation. \cite{andersnat07} showed that the non-linear feedback of dust onto
gas can lead to the rapid formation of gravitationally bound clumps of dust,
which subsequently form Ceres-size bodies. The dust particles, however, must
have already grown to some meters in size before this scenario can take
place. Another possibility for solving the formation problem is particle
trapping in gas pressure maxima \citep{bargesomm95,KlaHen97}. In gas pressure
bumps, relative particle velocities are substantially lower and, hence, the
collision between large bodies is more likely to lead to particle growth than
to particle disruption.

Interesting means of gas pressure maxima and dust particle trapping was
presented by \cite{kretke07}. This particle retention mechanism requires the
presence of an evaporation front, for example the snow line, which acts in the
following way. As we travel through the snow line in a direction away from the
central star, the dust-to-gas ratio increases suddenly. This jump in dust
density affects the strength of the magneto-rotational turbulence, since the
amount of free electrons in the disks strongly depends on the dust
density. With increasing dust density, the amount of turbulence in the disk
decreases \citep{sano00,ilg106}. We assume a constant mass accretion rate
throughout the disk. The gas surface density then has to be higher in low than
in high disk turbulent regions for the gas mass accretion to remain at a
similar value. The sudden increase in the dust density could therefore also
reproduce a sudden increase in gas densities. For certain accretion rates,
\cite{kretke07} found the occurrence of a local gas density bump in which
solid particles tend to accumulate. In their simulations, dust particle
retention reproduced very high surface dust densities of the order of several
$10^3$~g/cm$^2$, which again raises concern about gravitational instability.

However, even for a particle retention mechanism similar to that described
above, the growth of solid material towards larger sizes remains an open
issue. Gas pressure maxima may decrease relative radial particle velocities
triggering coagulation; the most severe reason for violent particle
fragmentation is not however radial drift but the turbulent nature of the
protostellar disk itself. Relative turbulent particle velocities are
approximately $10$~m/s \citep{ormel07,voelk80} and particle sticking at these
high speeds is extremely unlikely. Nevertheless, \cite{ciesla07} found that
particles can grow to several 10~m in radius if we consider layered MRI active
disks. Under certain circumstances, MRI is only active in the upper layers of
the disk, while the disk midplane is almost laminar. Since most of the larger
grains are located around the midplane where the disk is quiescent, dust
growth is not inhibited by the high speed collisions that produce particle
fragmentation.

We combine three ingredients for a planetesimal formation model. We consider
dust particle growth, particle fragmentation and radial motion \citep{brauer2}
around the snow line \citep{kretke07} in a layered MRI active protoplanetary
disk \citep{ciesla07}. The inclusion of a snow line in our simulations
provides a particle retention mechanism which almost erases the radial drift
velocities favouring particle growth. We adopt a layered MRI-driven disk to
circumvent particle fragmentation in the midplane of the disk due to turbulent
motions of the gas. We investigate whether solid particles can overcome the
fragmentation barrier and produce larger objects, which represent the possible
precursors to the planets. We study the influence of two parameters, namely
the gas accretion rate $\dot{M}_{\mathrm{acc}}$ and the critical threshold
velocity for fragmentation $v_{\mathrm{f}}$.

\section{Model}\label{model}

\begin{figure}
\begin{center}
\includegraphics[scale=0.48]{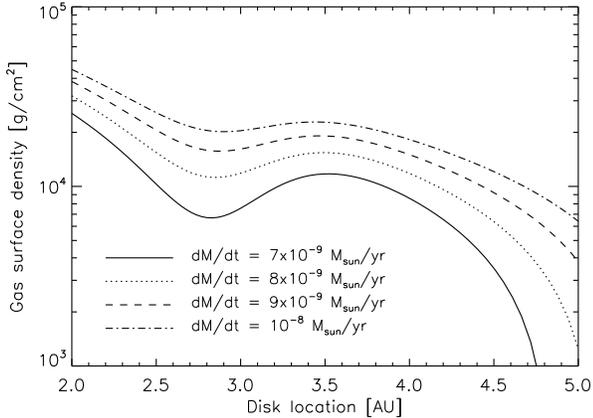}
\caption{The surface gas densities for different gas accretion rates
  which we will use in our simulations as calculated by
  \cite{kretke07} and discussed in Sec.~\ref{model}. The snow line is
  located at 3~AU.
 \label{sigma}}
\end{center}
\end{figure}

We assume a background gas disk which is in a steady state and, hence, does
not change with time. The calculation of the gas densities around the snow
line are described in detail in \cite{kretke07}. We adopt all of the parameter
values presented in that paper apart from the residual turbulence value around
the midplane $\alpha_0$, which we set to be $10^{-5}$ corresponding to
self-induced turbulence \citep{Wei79,WeiCuz93}. The water evaporation front,
which is the important element in our model, is located at 3~AU
\citep{lecar06}. The gas surface densities that we adopt in our simulations
are shown in Fig.~\ref{sigma} for different accretion rates
$\dot{M}_{\mathrm{acc}}$.

We focus on the dust component of the disk. We include the radial drift motion
of solids in our model. The maximum radial inward drift velocity is given by
\begin{equation}\label{vneq}
v_{\mathrm{N}}=\frac{\partial_rP_{\mathrm{g}}}{2\rho_{\mathrm{g}}\Omega_{\mathrm{k}}}\;.
\end{equation}
\begin{figure}
\begin{center}
\includegraphics[scale=0.48]{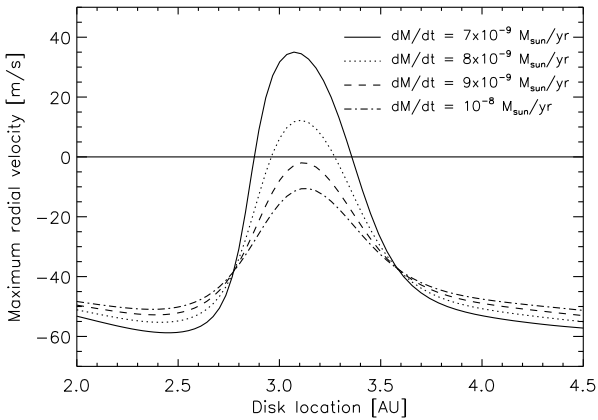}
\caption{Maximum radial drift velocity $v_{\mathrm{N}}$ for different
  gas accretion rates as discussed in Sec.~\ref{model}. The horizontal
  line indicates zero radial drift. Outward drift occurs for the two
  accretion rates $7\times 10^{-9}$ and $8\times
  10^{-9}$~M$_{\mathrm{sun}}$/yr. Hence, we expect dust grain
  retention for these two values. Note as well, that for these two
  accretion rates, there are two stagnation points in which the radial
  drift speed vanishes.
 \label{vn}}
\end{center}
\end{figure} 
For different accretion rates, this quantity is shown in Fig.~\ref{vn} as a
function of disk location. The equations for the gas pressure $P_{\mathrm{g}}$
and the Kepler frequency $\Omega_{\mathrm{k}}$ are given in
\cite{brauer2}. From this maximum radial drift velocity, we can calculate the
actual radial drift speed of a particle of a certain size and solve the
continuity equation in the radial direction for all dust particle
species. Radial diffusion due to turbulent mixing is is addition included in
the model.

Fig.~\ref{vn} shows, that for sufficiently small accretion rates, grain
retention will occur around the snow line close to 3~AU inside the disk. For
dust particle growth, it is even more important to note that there are two
radial drift stagnation points, i.e. disk locations at which the radial drift
completely vanishes. Relative particle speeds at these radii are too low to
induce destructive particle collisions.  While solid particles move away from
the inner diverging point at $\sim 2.9$~AU, the dust tends to accumulate in
the outer converging point at $\sim 3.3$~AU. It is also interesting to note
here, that the azimuthal gas flow becomes Keplerian at these stagnation
points. The differential azimuthal speed between the dust and gas vanishes,
which implies that self-induced turbulence is stablized regardless on how thin
the dust layer becomes.

We consider that the disk has a vertical structure. The vertical dust
distribution is determined by turbulent diffusion and particle settling
towards the midplane \citep{dms95,schr04}. We solve the
sedimentation/diffusion equation with a semi-analytical model, assuming that
the particle distribution is in equilibrium between particle settling and
turbulent mixing. We consider a dead zone around the midplane in which almost
no turbulence is present. In this region, we adopt an $\alpha$-value of
$10^{-5}$. In the MRI active regions on the surface of the disk, we assume an
$\alpha$ parameter of $1.8\times 10^{-2}$ \citep{sano98}. The amount of gas
and dust mass that is MRI active/inactive is provided in
\cite{kretke07}. Little turbulence around the midplane involves small relative
turbulent velocities \citep{ormel07}. Hence, in the radial drift stagnation
points, {\it all} relative velocities are low and we may expect dust particles
to grow to larger sizes and break through the fragmentation barrier in the
midplane of the disk.

At every point in the disk, we allow dust particles to coagulate and grow to
larger sizes or suffer fragmentation due to high speed collisions. The
parameter that distinguishes between these two cases is the collision velocity
of the particles. If this quantity is higher than a critical threshold
velocity $v_{\mathrm{f}}$, fragmentation occurs. Apart from the accretion rate
$\dot{M}_{\mathrm{acc}}$, the threshold fragmentation velocity is the second
parameter that we consider. In the simulations, we assume five different
sources of relative particle velocities in the disk that generate dust growth,
namely Brownian motion, differential settling, relative turbulent velocities,
and relative radial and azimuthal velocities. All of these sources are
explained extensively in \cite{brauer2}, which also describes the numerical
schemes adopted to solve the coagulation equation (i.e. Smoluchowski
equation), the outcome of fragmentation, and the effect of cratering that is
also included in the model.

In the final simulations completed by \cite{brauer2}, particles were always
smaller than the mean free path of the gas at any point in the disk. This is
not the case here due to the very high gas densities. This implies that we
must consider a different drag force regime, namely the Stokes regime instead
of the Epstein regime. We implemented this regime into our model to account
for the high gas densities \citep{Wei77}.

\section{Results}\label{res}

\begin{figure}
\begin{center}
\includegraphics[scale=0.48]{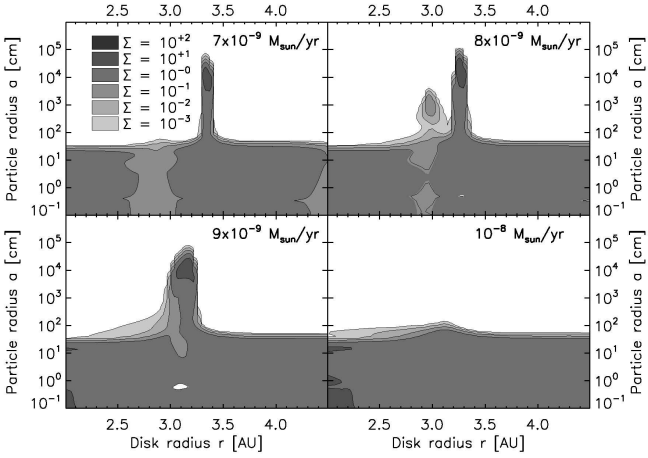}
\caption{The particle distribution after 1800~yrs of disk evolution as
  discussed in Sec.~\ref{res}. Shown is a contour plot of the surface dust
  density as a function of disk location and particle radius for four
  different accretion rates. The figure indicates that particles break through
  the fragmentation barrier if the accretion rate is not too high. The
  critical fragmentation velocity is 10~m/s in this simulation.
 \label{cmdot}}
\end{center}
\end{figure}

Fig.~\ref{cmdot} shows the particle distribution after 1800~yrs of disk
evolution for different accretion rates. In this simulation, we adopted a
critical threshold fragmentation velocity of 10~m/s.
This plot indicates that particles can grow to some $10^2$~m in size around
the ice evaporation front subject to the condition that the gas accretion rate
is not too high. It also shows that fragmentation inhibits particle growth
towards m-sized particles in other disk regions, for example around 4~AU.  To
unveil the importance of collective effects at the snow line, we estimate the
dust-to-gas ratio in the midplane of the disk. The vertically integrated ratio
in the case of $\dot{M}=9\times 10^{-9}$~$M_{\mathrm{sun}}$/yr is of the order
$\epsilon_0=10/10^4=10^{-3}$, as inferred from Figs.~\ref{sigma}
and~\ref{cmdot}. The dust-to-gas ratio in the midplane is then at least
$\epsilon_{\mathrm{mid}}\approx\epsilon_0\sqrt{\mathrm{St}/\alpha}\approx
10^{-3}\sqrt{10^3/10^{-5}}=10>1$, which means that collective effects play a
non-negligible role. Since collective effects strongly influence the radial
drift behaviour of the dust, further investigation of this issue is certainly
required.

The accretion rate of a protoplanetary disk decreases with time. Therefore,
Fig.~\ref{cmdot} illustrates the advantage of our planetesimal formation
mechanism at different stages of disk evolution. In the early stages, the gas
surface densities are not significantly affected by the sharp decrease in
dust-to-gas ratio at the snow line because the accretion rate is too high.
Hence, we do not find that particles grow to become very large objects in this
case. With decreasing accretion rate, the surface gas density becomes more and
more affected by the snow line producing the formation of large boulders as
shown in Fig.~\ref{cmdot}. At this stage of disk evolution, planetesimal
formation could occur. At later stages, the surface gas densities will have
declined towards a level at which the entire disk becomes MRI active and the
mechanism is unable to operate.

\begin{figure}
\begin{center}
\includegraphics[scale=0.48]{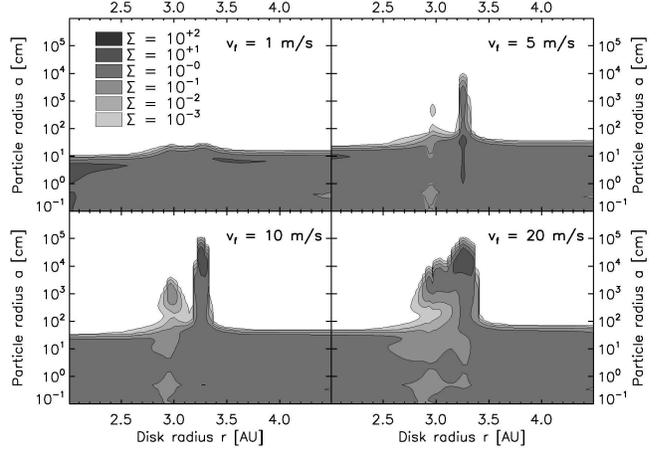}
\caption{As Fig.~\ref{cmdot}, but now for four different critical theshold
  fragmentation velocities. The figure indicates that dust particles can break
  through the fragmentation barrier if the theshold fragmentation velocity is
  at least 5~m/s. In this simulation, we adopted an accretion rate of
  $\dot{M}_{\mathrm{acc}}=8\times 10^{-9}$~$M_{\mathrm{sun}}$/yr.
 \label{cvf}}
\end{center}
\end{figure}

The dust particle distribution after 1800~yrs, but now for different
criticial fragmentation velocities, is shown in Fig.~\ref{cvf}. In
this calculation, we considered an accretion rate of
$\dot{M}_{\mathrm{acc}}=8\times 10^{-9}$~$M_{\mathrm{sun}}$/yr.
This plot indicates that dust particles can overcome the fragmentation barrier
and grow to nearly km-size if the critical threshold velocity is at least
5~m/s. For the case $v_{\mathrm{f}}=1$~m/s, we do not find particles that are
able to grow to larger than a meter in radius. This is due to the fact that
relative turbulent velocities in our quiescent midplane are still of the order
of several m/s. Even with a low residual turbulent $\alpha_0$-value of
$10^{-6}$, relative particle velocities would hardly decrease to below this
value.

The residual turbulent $\alpha_0$-value around the midplane can be shifted to
higher values due to the occurrence of turbulent convection. \cite{KlaHen97}
suggest turbulent gas speeds of $\sim 0.02 c_{\mathrm{s}}$, which correspond
to velocities of $\sim 16$~m/s at 1~AU and an $\alpha$-value of $\sim
10^{-4}$. Hence, if convection is a driving source of turbulence in the disk
then the results of this Letter might change significantly. \cite{turner07}
showed that free charges can be mixed into the disk interior, producing a
slight coupling between the midplane gas and the magnetic fields. This may
generate an active turbulent midlane layer, which could trigger particle
fragmentation. Further investigation of the influence of each of these effects
is imperative.

\section{Summary and conclusions}

Even though particle fragmentation by high speed collisions is a severe
obstacle to the formation of planetesimals, we have shown that dust particle
growth to the size of enormous boulders is possible under certain
circumstances. If we consider dust particle coagulation in the presence of an
evaporation front, solid particles grow to almost km-sizes within only a few
thousand years. The protoplanetary disk must also contain a dead zone in which
turbulence -- the main source of relative particle velocities -- is almost
entirely absent. Our conclusion is that planetesimal formation due to dust
particle agglomeration is a possible mechanism to form large bodies in
protoplanetary disks.

\section*{Acknoledgement}

We wish to thank A.~Johansen and H.~Klahr for helpful discussions. We also
thank the anonymous referee for useful comments that helped us to improve this
Letter.

\bibliographystyle{aa} \bibliography{refsv2.bib}

\end{document}